\documentclass[journal=jpccck,manuscript=article]{achemso}

\usepackage[version=3]{mhchem} 
\usepackage[T1]{fontenc}       

\usepackage{subfigure}
\usepackage{graphicx}
\usepackage{dcolumn}
\usepackage{bm}

\usepackage[dvips]{color}

\author{C. Garc\'{\i}a Fern\'{a}ndez}
\affiliation{Donostia International Physics Center (DIPC), Paseo Manuel de Lardizabal 4, 20018 Donostia-San Sebasti\'an,Spain}
\email{dr. carlos.garcia@dipc.org}

\author{P. Abufager}
\affiliation{Instituto de F\'{i}sica de Rosario, Consejo Nacional de Investigaciones
Cient\'{i}ficas y T\'ecnicas (CONICET) and Universidad Nacional de Rosario,
Bv. 27 de Febrero 210 bis (2000) Rosario, Argentina}

\author{N. Lorente}
\affiliation{Centro de F{\'{\i}}sica de Materiales
CFM/MPC (CSIC-UPV/EHU), Paseo Manuel de Lardizabal 5, 20018 Donostia-San Sebasti\'an, Spain}
\affiliation{Donostia International Physics Center (DIPC), Paseo Manuel de Lardizabal 4, 20018 Donostia-San Sebasti\'an,Spain}

\title[An \textsf{achemso} demo]{Transport properties of Co in Cu(100) from first principles.} 

\begin{document}

\begin{abstract}
The electronic transport properties of a point-contact system formed by
a single Co atom adsorbed on Cu (100) and contacted by a  copper tip
is evaluated in the presence of intra-atomic Coulomb interactions and
spin-orbit coupling.  The calculations are performed using equilibrium
Green's functions evaluated within density functional theory completed
with a Hubbard $U$ term and spin-orbit interaction, as implemented in
the Gollum package. We show that the contribution to the transmission
between electrodes of spin-flip components is negative and scaling
as $\lambda^2/\Gamma^2$ where $\lambda$ is the SOC and $\Gamma$ the
Co atom-electrode coupling.  Hence, due to this unfavorable ratio,
SOC effects in transport in this system are small. However, we show
that the spin-flip transmission component can increase by two orders of
magnitude depending on the value of the Hubbard $U$ term.  These effects
are particularly important in the contact regime because of the prevalence
of $d$-electron transport, while in the tunneling regime, transport
is controlled by the $sp$-electron transmission and results are less
dependent on the values of $U$
and SOC. Using our electronic structure and the elastic transmission
calculations, we discuss the effect of $U$ and SOC on the well-known
Kondo effect of this system.  
\end{abstract}

\section{Introduction}
\label{intro}

The study of single magnetic atoms on non-magnetic metals has
become a reality thanks to the advent of local scanning probe
microscopies~\cite{Gauyacq_2012,Markus_2017}. This
is a priviledged situation in which precise measurements can
be performed on a very controlled environment.  Many
theoretical works have been recently undertaken to
quantitatively evaluate the properties revealed
in these experiments~\cite{Gauyacq_2012,Markus_2017,Delgado_2017}. Among these properties, 
Kondo physics~\cite{Kondo,Hewson} is currently the object of much interest~\cite{Berndt,Crommie,Knorr_2002,Neel_2007,Markus_2009,Vitali_2008,Choi_2012,Choi_2016}.

Here, we take one of these almost-ideal
systems and perform  calculations to unravel
the electronic structure and its effect on
the electronic transport revealed by
the experiments. The system is a single Co impurity
adsorbed on a Cu (100) surface that
is contacted by a scanning tunneling microscope (STM)
tip~\cite{Knorr_2002,Neel_2007,Vitali_2008,Choi_2012}.

Many interesting effects have been found
for this system. Polok and co-workers~\cite{Polok_2011} found
that electron transport qualitatively changed
with the tip-adatom distance. When the tip was
far from the substrate, transport took place through
the $sp$-induced electronic structure of the adatom. When
the tip-apex atom approached until reaching covalent-bond
distances, the electronic transmission involved the $d$-system. More
curiously, the effect of the tip was a reordering of the electronic
structure, changing the system's properties depending on
the tip-surface distance. Unfortunately, these calculations
did not address the very interesting Kondo physics
experimentally revealed~\cite{Vitali_2008,Choi_2012}.

Direct calculations of the Kondo physics of Co on Cu (100)
showed that the dynamical correlation processes where
basically controlled by the $d_{z^2}$ orbital of Co~\cite{Baruselli_2015,Jacob_2015,Frank_2015}. Hence,
the problem seemed to be greatly simplified by just considering
the behavior of the $d_{z^2}$ orbital
as the tip-surface distance changed~\cite{DJ_Paula}.
However, all of these calculations where based on
the results of plain density functional theory (DFT), which
are known to underestimate the magnetic properties
of adsorbed impurities. Particularly, two important
ingredients present in modern calculations were missing
from the above studies: the intra-atomic Coulomb
interaction as given by the Hubbard $U$, and
the spin-orbit coupling (SOC) of the Co atom.

In this work, we consider the effect of both interactions
in the electronic structure of Co adatoms both
in the tunneling regime that corresponds well to
the Co / Cu (100) adsorbed system, and to
the contact regime where the tip creates a covalent
bond with the Co adatom becoming a  point-contact junction.
The paper contains a first section devoted to the methodology
 and setup of the calculations.
We first show the results of the electronic structure
and transmission calculations for different values of
the Hubbard $U$. Next, the paper considers the effect of 
SOC in the Co atom and its influence on the electronic
transmission. We analyze the electron transmission by
simplifying the problem to the $d$-manifold of the Co atom
and we rationalize the effect of SOC on the
electron transmission in terms of the strength of
the SOC  as compared to electronic coupling
the Co $d$ orbitals with the electrodes. We analyze the
consequences of our findings on the Kondo effect
and summarize the article.

\section{Computational methodology}
\label{method}
In this work, we depart from two of the reported  optimized
geometries of Choi et al.\cite{DJ_Paula} for an atomic junction formed by
an adsorbed Co atom on a Cu(100) surface and a copper-covered tip. The
two surfaces representing substrate and tip were modeled using a periodic slab
geometry with a 3$\times$3 surface unit cell, 6 layers for the surface
holding the Co atom and 5 layers for the tip electrode. 

\begin{figure}[h!]
\centering
\includegraphics[width=0.45\textwidth]{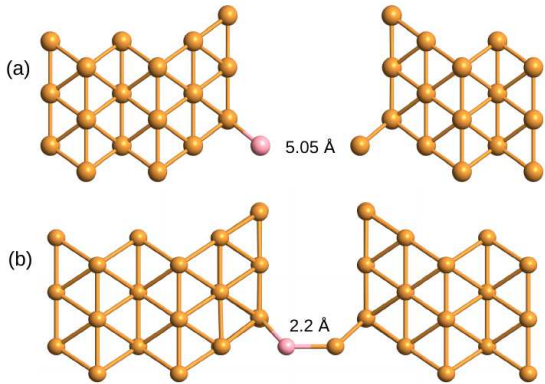}
\caption{
Atomic models with the two configurations used along
this work:(a) the tunneling regime where the distance
between the Co atom and the tip apex is
5.05~\AA~and (b) the contact regime where the same distance diminishes until
typical covalent-bonding distances, here 2.2~\AA. Periodic boundary conditions have been applied 
along the axes perpendicular to the transport direction.
}
\label{atomos}
\end{figure}

DFT calculations have been performed within the spin-polarized
generalized gradient approximation (GGA-PBE) \cite{PBE}. Troullier-Martins
full-relativistic pseudo-potentials \cite{martin}, an energy cutoff of
500 Ry and a 7$\times$7$\times$3 K-point mesh generated according to the
Monkhorst-Pack scheme, have been used in the Siesta code \cite{siesta}.
A double-$\zeta$ plus polarization (DZP) basis set was defined to
describe the Co and surface-atom electrons, while diffuse orbitals
were used to improve the surface electronic description.  Furthermore,
a single-$\zeta$ plus polarization (SZP) basis was set for the copper
electrodes.  Note that we employ a DZP basis set to describe the adsorbate
states in order to yield correct transmission functions \cite{Paula}.


 
Quantum transport computations were performed from first-principles
within the framework of the Landauer-Buttiker formalism.  Thus, the DFT
Hamiltonian and overlap matrices obtained with Siesta \cite{siesta} were analized in a
post-processing step with the Gollum package~\cite{gollum}.  This code
is based on equilibrium transport theory and by carefully setting
electrodes, branches and the central scattering region~\cite{manual}, the
transmision coefficients can be computed without performing independent
selfconsistent calculations of the density matrix. This approach results
in considerable savings of time and computational resources. Furthermore,
one of the attractive features of Gollum is its functionality to compute
spin transport in systems with spin-orbit interactions.
Comparison with previous calculations~\cite{DJ_Paula} using self-consistent non-equilibrium
Green's function calculations shows that both calculations
agree within the available numerical precision.

\section{Results and discussion}
\subsection{Electronic structure and transport}
\label{structure}

The goal of the present section is to explore the sensitivity of the
electronic and transport properties of the Cu-Co-Cu junction with respect
to two distinct interactions: the Coulomb on-site repulsion, or Hubbard
$U$, on Co-$d$ orbitals and the spin-orbit coupling (SOC).

We first analize the effect of the $U$ parameter on the electronic
and transport properties of the system comparing with the results of
standard DFT (for the PBE exchange-and-correlation functional) in the
tunnel Fig.~\ref{tunnel} and contact regimes Fig.~\ref{contact}.
Next, we include spin-orbit coupling and evaluate the same
properties. We simplify the calculations to include just the
$d$-electrons of the Co atom and rationalize our
findings at the end of this section.

\subsection{Effect of $U$ on the electronic and transport properties}

 \begin{figure*}[t] 
\centering
\includegraphics[width=0.8\textwidth]{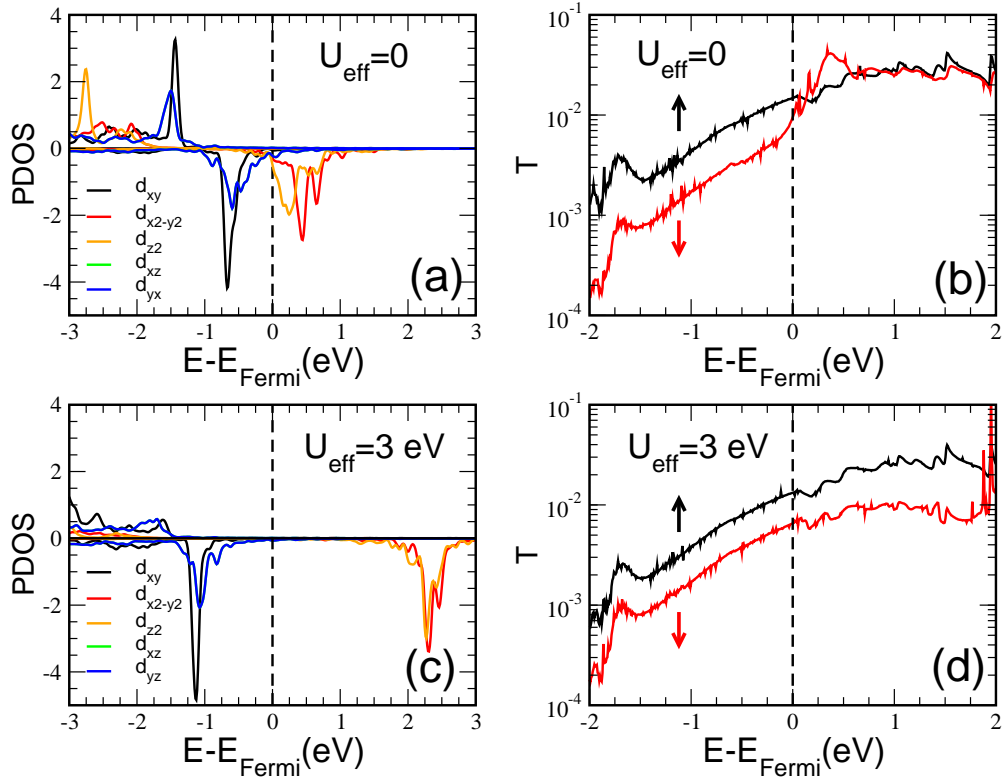}
\caption{Results in the tunneling configuration for: (a,c) Density of states projected onto Co-$d$  atomic orbitals  (PDOS)
and (b,d) electron transmissions (T)  between
electrodes as a function of electron energy referred to the Fermi energy ($E-E_{Fermi}$). 
No spin-orbit coupling is included and the graphs are divided in majority (positive PDOS and up arrow for T)
and minority (negative PDOS and down arrow for T) spins.
The Hubbard-U of the Co $d$-manifold used in the GGA+U scheme are $U_{eff} = 0$ eV in the upper graphs (a,b), and $U_{eff}$ = 3 eV in 
the lower ones, (c,d).}
\label{tunnel}
\end{figure*}

\begin{figure*}[t] 
\centering
\includegraphics[width=0.8\textwidth]{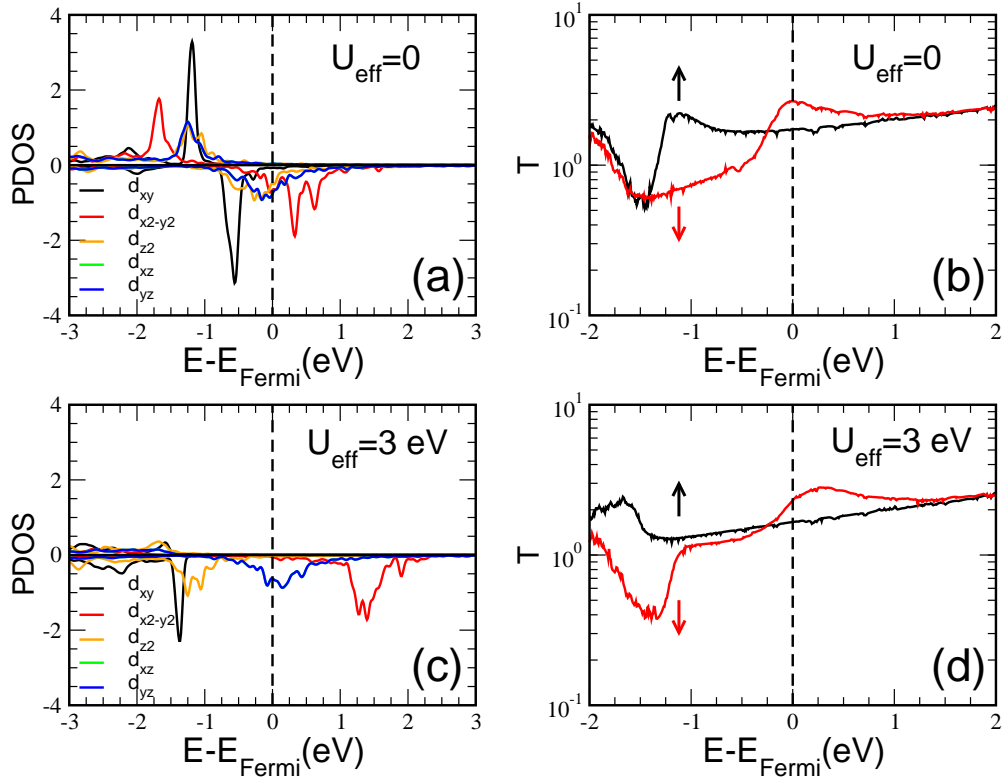}
\caption{Results in the contact configurations for: (a,c) PDOS and (b,d) electron transmissions (T)
with the same parameters and conventions as in Fig.~\ref{tunnel}.}
\label{contact}
\end{figure*}

Fig.~\ref{tunnel}(a) and ~\ref{contact}(a) shows the PDOS projected onto
Co-$d$ atomic orbitals, computed in this study without the inclusion of the Coulomb on-site repulsion, 
in the tunneling and contact configurations respectively.

When the tip is far from the cobalt atom, the two singly occupied
$d_{z^2}$,$d_{x^2-y^2}$ magnetic orbitals are clearly shown as unoccupied
states, in good agreement with the results by Polok \textit{et al.}~\cite{Polok_2011} and
Baruselli \textit{et al.}~\cite{Baruselli_2015}.  

However, the detailed electronic structure of the Co adatom
changes in the contact region. There is a re-ordering of the minority
spin $d$-states and contributions from $d_{z^2}$, $d_{xz}$ and $d_{yz}$
orbitals become important at the Fermi level~\cite{Polok_2011,DJ_Paula}.  In spite of the changes
observed in the electronic structure, the overall magnetic properties
slightly change since the Co adatom can be described as in a 3$d^8$
(S = 1) configuration in both regimes~\cite{Polok_2011,Baruselli_2015,DJ_Paula,Surer_2012,Jacob_2015,Frank_2015}.

The re-ordering of $d$ levels in contact induces changes in the transport
properties, as was previously reported \cite{Polok_2011,DJ_Paula}.  In the tunneling
region, we find that transport Fig. ~\ref{tunnel} (b) basically takes place through the
majority spin $sp$ electrons of the Co atom. 

In the contact regime, however, the $d$-electron contribution to the transmission at the Fermi level for the
minority spin  highly increases and the transport is governed by the minority spin channel. 
Indeed, the spin polarization defined as $P = (T_\uparrow(E_F) - T_\downarrow(E_F))/(T_\uparrow(E_F)
+ T_\downarrow(E_F)) $, where $T_\sigma(E_F)$ is the transmission per spin $\sigma$
at the Fermi energy, changes its sign when going from tunneling to contact.


As reported before\cite{Polok_2011,DJ_Paula}, our present calculations
confirms the
previous picture where  conduction  takes place through the $sp$ electrons
of the Co adatom in tunneling while Co $d$-orbitals  dominate the transport
in the contact regime.



The Coulomb on-site repulsion on Co $d$-orbitals is an indispensable
component in the above scenario.  Since such interaction could modify
the picture if the partially occupied $d_{z^2}$, $d_{xz}$ and $d_{yz}$
orbitals are pulled down with respect to unoccupied orbitals when
$U$ is increased.  Here, we
explore such an issue by performing LDA+U calculations  through the
simplified rotationally invariant formulation of Dudarev et
al \cite{Dudarev_1998}, with an effective $U_{eff}=U-J$
that includes the effect of the Fock exchange interaction, $J$. We use the
implementation of the Siesta code~\cite{Daniel}.

\begin{figure}[h!]
\centering
\includegraphics[width=0.4\textwidth]{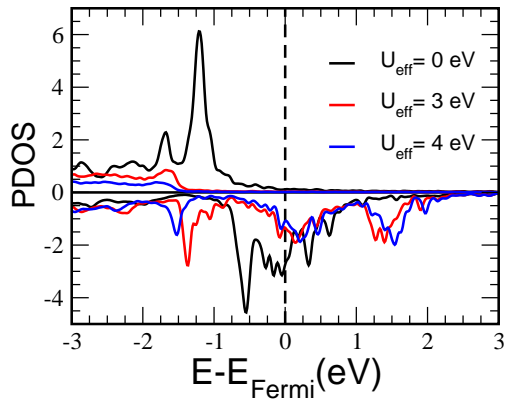}
\caption{Spin-polarized PDOS projected onto Co(3$d$) atomic 
orbitals at contact for  U$_{eff}=U-J = 0, 3, 4$ eV 
}\label{conv_U}
\end{figure}

As depicted in Fig.~\ref{tunnel}(c) ($U_{eff}=3$ eV) the overall properties for the tunnel
regime slightly change with respect to the PDOS computed at the DFT level
(Fig.~\ref{tunnel}(a)), being the main quantitative feature related to
an increase of the energy separation among $d$-levels.  However, when the tip is
close to the surface Fig.~\ref{contact}(c), besides the separation
among levels, the minority $d_{z^2}$ is shifted to lower energies,
leaving the $d_{xz}$ and $d_{yz}$ as the only orbital contributions
around the Fermi energy.  The increase of the separation among levels
and the decrease of the minority channel contribution at the Fermi energy when
including the Coulomb repulsion can be clearly seen in Fig.~\ref{conv_U}
depicting the total Co 3$d$ PDOS for three different values of the
effective $U_{eff}$.

In transport, a large Co $sp$ contribution remains in the
transmission at the Fermi level.  As a result, good agreement is
found between $U_{eff}=0$ eV and $U_{eff}=3$ eV calculations as can
be seen from Figs. ~\ref{tunnel} (b) and (d).  When the tip is near the
surface adatom, two differences are clearly seen between  $U_{eff}=3$
eV (Fig.~\ref{contact}(d)) and  $U_{eff}=0$ eV (Fig.~\ref{contact}(b)).
There is a shift in the energy scale, related to
the new energy position of the Co orbitals. The second difference is
an increase of $4\%$ and $15\%$ of $T(E_F)$ with respect to the
$U_{eff}=0$ values computed  for the majority and minority
spin channels, respectively.  This difference can be traced back to a change
in the $d_{z^2}$ orbital energy.  

In spite of these differences, the overall scenario of
Refs.~\cite{Polok_2011,DJ_Paula} based on the leading transmission through
the minority spin channel governed by Co-$d$ orbitals  remains unchanged.
Therefore, we can conclude that the Coulomb on-site repulsion on Co-$d$
orbitals does not have a strong effect on the transport properties of
the Co junction.

\subsection{Electronic transport in the presence of spin-orbit coupling}
\label{spin-orbit}

\begin{figure}[h]
\centering
\includegraphics[width=0.4\textwidth]{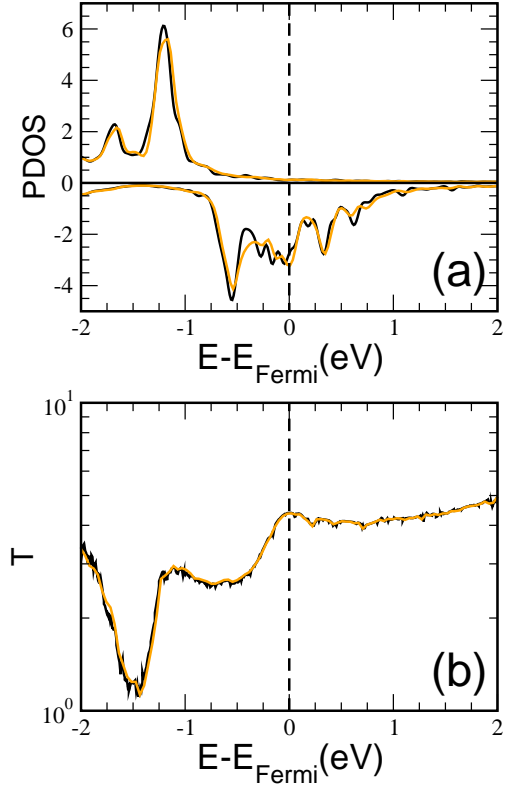}
\caption{Total (a) PDOS projected onto Co(3d) atomic orbitals and 
(b) transmission at contact with (orange line) and without 
(black line) taking into account spin-orbit interactions and $U_{eff}=0$.}\label{SOC} 
\end{figure}

We include spin-orbit coupling (SOC) in the  DFT equations
following the implementations in Siesta~\cite{siesta} and
Gollum~\cite{gollum}. 
The PDOS, Fig.~\ref{SOC} (a) ($U_{eff}=0$), reveals that
the effect of SOC on the electronic structure of
the Co-adatom is negligible within the accuracy of DFT calculations.  
This translates into the calculations of the electron transmission,
showing that the presence of SOC produces no effect.
Figure~\ref{SOC}(b) compares the
transmission with and without SOC. Both curves
agree. However, the plotted total transmission is
the sum of four terms. Mainly the direct non-spin-flip transmissions
and the transmissions where the electron spin changes between
electrodes. In the next section, we study the 
 spin-flip contributions to the transmission and their sensibility to the Coulomb on-site repulsion.

\subsection{Spin-orbit assisted spin-flip scattering}

The present problem consists in a single center where the SOC is localized. We
use the electronic structure computed above (with and without the inclusion of $U_{eff}$) to compute
the transmission by just considering the transmission through the $d$-orbitals
that are the ones containing sizeable SOC contributions.
This approximation is particularly well fitted to the contact regime.
In Ref.~\cite{DJ_Paula} three channels of strong $d$ character are
shown to dominate the transmission at contact while the $sp$ channels
dominate transport in the tunneling regime.

The transmission between electrodes for an electron injected at energy $E$ through the $d$-orbitals
of the Co atoms is:
\begin{equation}
T(E)=\sum_{i,j,k,l}\Gamma^L_{i,j} G_{j,k}^r \Gamma^R_{k,l} G_{l,i}^a.
\label{Landauer}
\end{equation}
Where ${i,j,k,l}$ are indices over the spin orbitals of the $d$-electron manifold. The
Green's functions $G_{j,k}^r$ and $ G_{l,i}^a$ are the retarded and advance resolvents
of the atomic Hamiltonian, $\hat{H}$, in contact with the two electrodes, expressed again in the $d$-electron spin orbitals:
\begin{equation}
G_{j,k}^{r (a)}=\langle i | [E \hat{1} - \hat{H} -\hat{\Sigma}^{r (a)}]^{-1}  |j \rangle.
\label{green}
\end{equation}
The identity operator, $\hat{1}$, becomes a matrix of the dimension of the $d$ manifold as well
as the retarded (advanced) self-energy, $\hat{\Sigma}^{r (a)}$.
The imaginary part of the self-energy is actually related to $\Gamma$ of each electrode by 
(here $i$ is the imaginary unit):
\begin{equation}
\hat{\Gamma}=i\hat{\Sigma}^{r}-i\hat{\Sigma}^{a}.
\label{self-energy}
\end{equation}
The total self-energy is the sum of self-energies due to each electrode.

In the spirit of the above calculations, we use Kohn-Sham orbitals, and
the problem becomes a one-electron transport problem. From this analysis
we find that for $E$ between $-2$ and $2$ eV transport at contact is
dominated by three $d$ orbitals in good agreement with previous results,
Ref.~\cite{DJ_Paula}. Prior to switching on the SOC, we can identify
the three spin orbitals as the minority spin degenerated $d_{xz}$ and
$d_{yz}$ with some further contribution from the $d_{z^2}$ orbital.
This is particularly true for the $U_{eff}=0$ cases, Figures~\ref{contact}
(a) and (b). The minority spin peak that dominates the transmission
at the Fermi energy is clearly a contribution of the just mentioned
orbitals. However, for $U_{eff}=3$ eV, the $d_{z^2}$ orbital shifts down in
energy and the transmission at the Fermi energy is controlled by the
minority-spin degenerated $d_{xz}$ and $d_{yz}$.  The comparison of
Figs.~\ref{contact} (a) and (c) show the clear reduction of the weight
of $d_{z^2}$-type electronic structure at the Fermi energy. This is
concomitant with the appearance of a sharp minimum at $\sim-1.5$~eV in
Fig.~\ref{contact} (d). Our calculations using the transmission through
$d$ orbitals, Eq.~(\ref{Landauer}), show that its origin is a sizeable
interference term $\Gamma_{z^2,xy}$ due to the mixing of orbitals by
the Cu $d$-band that starts at $\sim-1.8$~eV.

The SOC is included in Hamiltonian, $\hat{H}$, restricted to the $d$-electron
subspace. This is particularly simple to do in the
Cartesian representation of $d$ electrons.
We follow Ref.~\cite{Dai_2008}, where all matrix elements are carefully
written. We write the SOC contribution to the Hamiltonian as:
\begin{eqnarray}
\hat{H}_{SOC} &=& \lambda (\hat{L}_z
\hat{S}_z+\hat{L}_x \hat{S}_x+\hat{L}_y \hat{S}_y) \nonumber \\
&=&
\lambda (\hat{L}_z
\hat{S}_z+ \frac{1}{2} [\hat{L}_+ \hat{S}_-+\hat{L}_- \hat{S}_+]).
\label{VSOC}
\end{eqnarray}
While the first term connects orbitals with the same $|m|$
and spin, where
$m$
is the eigenvalue of $\hat{L}_z$, the second term leads to
spin-flips connecting spin-orbitals with different spins and orbitals
of $|m\pm 1|$. If we consider the matrix elements in Cartesian
terms, we also remark that while the matrix elements of
$\hat{L}_z\hat{S}_z$ and $\hat{L}_x \hat{S}_x$
are purely imaginary, $\hat{L}_y \hat{S}_y$ are purely real.
All diagonal matrix elements are zero because in Cartesian orbitals
the average angular momentum is zero.\\
\\

%
%
\begin{figure}[h]
\centering
\includegraphics[width=0.4\textwidth]{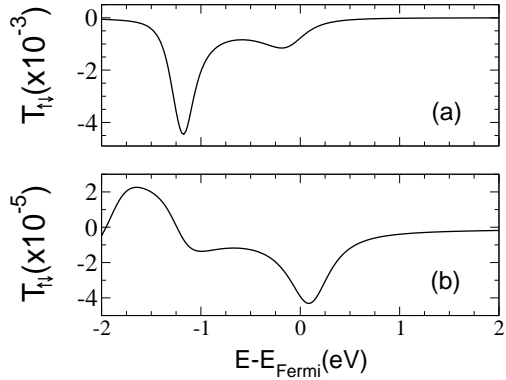}
\caption{Spin-flip component of the transmission for (a) $U_{eff}=0$ and (b) $U_{eff}=3$ eV in the
Co $d$-manifold. 
There is a factor-of-100 difference in the transmission between (a) and (b) marked
in the $y$-axis label.
}\label{TSOC}
\end{figure}

It is straightforward to build the new Green's function by inverting
the old Hamiltonian with the above additional term, Eq.~(\ref{VSOC}).
The value we took for $\lambda$ was the one of Co (I) because DFT yields
a $3d^8$ state for Co ($S=1$) in the contact configuration~\cite{DJ_Paula}.
Using the values of Ref.~\cite{Dai_2005} we have that $\langle \xi \rangle=455$ cm$^{-1}$,
and
\begin{equation}
\lambda=\frac{\langle \xi \rangle}{2 S}=0.0284~\mbox{eV}.
\label{lambda}
\end{equation}

We retrieve our calculations using DFT with SOC for the
transmission function, Fig.~\ref{SOC} (b), with
the message that $\lambda$ is so small
that the effect on the transmission is negligible. 
As we will show in the next section, SOC effects
will be noticeable as soon as the ratio $(\lambda/\Gamma)^2$
is not small, where $\Gamma$ is the width of the
$d$ levels given by Eq.~(\ref{self-energy}).

Since the spin of
the electron in the electrodes is a good quantum number, we can
study the transmission of each spin. From Eq.~(\ref{Landauer}), we
single out the spins and follow the transmission of each spin. We
see that the transmission is a $2\times 2$ matrix in spin
due to the combinations of entering with either up or down spins
and exiting with up and down including cross terms where the
spin flips due to the presence of SOC at the Co atom. Let
us analyze the spin-flip term:
\begin{equation}
T_{\uparrow,\downarrow}(E)=Tr[\Gamma^L_{\uparrow} G_{\uparrow,\downarrow}^r \Gamma^R_{\downarrow} G_{\downarrow,\uparrow}^a].
\label{spinflip}
\end{equation}
$Tr$ stands for trace over $d$ orbitals and there are three matrix products because both $\Gamma$'s and $G$'s are matrices on $d$ orbitals.

Figure~\ref{TSOC} shows the results for the spin-flip transmission,
$T_{\uparrow,\downarrow}(E)$ of Eq.~(\ref{spinflip}). This contribution is
negative, leading to the decrease of electron transmission in the system
and to the increase of electron backscattering. Despite their small value
stemming from the smallness of $\lambda/\Gamma$, we see fundamental
differences between the two plotted cases. Figure~\ref{TSOC}(a)
is the $U_{eff}=0$ case while Fig.~\ref{TSOC}(b) is the $U_{eff}=3$ eV one. The
difference of two orders of magnitude between the two cases comes
from the different electronic structure. In the $U_{eff}=0$ case, the three
orbitals $d_{z^2}$, $d_{xz}$ and $d_{yz}$ are close in energy about
the Fermi level. These three orbitals have non-zero spin-orbit matrix
elements connecting them because of the above spin-flip rule. Namely,
the flipping of an electron leads to the change of $|m|$ where $m$ is
the third component of the angular momentum of the spherical harmonics
entering the orbital. Spin-flip then involves matrix elements changing
$|m|$ by one, which is the case between the degenerated $d_{xz}$ and
$d_{yz}$ with the $m=0$, $d_{z^2}$ orbital.  When $U_{eff}=3$ eV, the $d_{z^2}$
orbital moves by more than 1.5 eV away from the $d_{xz}$ and $d_{yz}$
orbitals, quenching the spin-flip probabilities.

\subsection{Simplified spin-orbit  transmission}

It is interesting to simplify the above treatment to enhance our
insight into the spin-orbit-induced spin flip. Let us
assume a single $d$ orbital. In this case the spin-orbit
contribution to the Hamiltonian becomes~\cite{Evers_2015}:
\begin{equation}
\hat{H}_{SOC} = i \lambda (\hat{d}^\dagger_\uparrow \hat{d}_\downarrow - \hat{d}^\dagger_\downarrow \hat{d}_\uparrow).
\label{simplify}
\end{equation}
This is obviously Hermitian, and the matrix element is  $i \lambda$, purely imaginary.

We adopt the broken-symmetry description of DFT, then, the atomic level becomes $\epsilon_\uparrow$ and $\epsilon_\downarrow=\epsilon_\uparrow+U$ with $U$
the Hubbard charging energy. Within the wide-band approximation,
the Green's function of the orbital in contact with two electrodes is:
\begin{equation}
G (E)=\begin{pmatrix}
E-\epsilon_{\uparrow}+i \frac{\Gamma^L_\uparrow+\Gamma^R_\uparrow}{2} & -i \lambda \\
i \lambda &  E-\epsilon_{\downarrow}+i \frac{\Gamma^L_\downarrow+\Gamma^R_\downarrow}{2} 
\end{pmatrix}^{\!\!-1}
\label{Gs}
\end{equation}
with obvious notations for the self-energies of the level due to the left and right electrodes
(real parts are strictly zero in the wide-band approximation) for each spin.
Replacing these quantities in Eq.~(\ref{Landauer}), we obtain for the direct terms:
\begin{equation}
T_{\uparrow,\uparrow}(E)=\frac{\Gamma^L_\uparrow \Gamma^R_\uparrow \; |E-\epsilon_{\downarrow}+i \frac{\Gamma^L_\downarrow+\Gamma^R_\downarrow}{2}|^2
}{|(E-\epsilon_{\uparrow}+i \frac{\Gamma^L_\uparrow+\Gamma^R_\uparrow}{2})(E-\epsilon_{\downarrow}+i \frac{\Gamma^L_\downarrow+\Gamma^R_\downarrow}{2})-\lambda^2|^2}
\label{tuu}
\end{equation}
In the limit $\lambda \rightarrow 0$ we retrieve the usual result:
\begin{equation}
T_{\uparrow,\uparrow} (E)=\frac{\Gamma^L_\uparrow \Gamma^R_\uparrow}{|E-\epsilon_{\uparrow}+i \frac{\Gamma^L_\uparrow+\Gamma^R_\uparrow}{2}|^2}.
\end{equation}

The spin-flip term is proportional to $\lambda^2$:
\begin{equation}
T_{\uparrow,\downarrow} (E)=\frac{\Gamma^L_\uparrow \Gamma^R_\downarrow \, (i \lambda)^2}{|(E-\epsilon_{\uparrow}+i \frac{\Gamma^L_\uparrow+\Gamma^R_\uparrow}{2})(E-\epsilon_{\downarrow}+i \frac{\Gamma^L_\downarrow+\Gamma^R_\downarrow}{2})-\lambda^2|^2}.
\label{tud}
\end{equation}
In the limit of large $\Gamma^R\sim \Gamma^L\sim \Gamma \gg \epsilon_\uparrow$ we see that the spin-flip contribution
to the transmission becomes a negative quantity quadratic on the $\lambda$ to $\Gamma$ ratio:
\begin{equation}
T_{\uparrow,\downarrow} (E)\sim - \frac{\lambda^2}{\Gamma^2}.
\label{tudlimit}
\end{equation}
This sets a scale for the values of $\lambda$ that yield sizeable spin-flip terms in electron transport. Typically $\Gamma$ is about a few hundred meV. If $\lambda$ is in the tens of meV ($3d$ transition metals) the spin-flip terms will be negligible in transport
for a single scattering center. However, heavy elements will produce important spin-flips in point contacts.

\subsection{Kondo effect}

The computed electronic structure has direct bearings on the
Kondo effect that a Co impurity displays in contact with copper
electrodes~\cite{Baruselli_2015,Jacob_2015,Frank_2015}. The very
different electronic properties of the calculations of Figs.~\ref{tunnel}
(a) and (c) depending on the value of the Hubbard $U$, will
change the interpretation of this Kondo effect. Indeed, for $U_{eff}=0$,
Fig.~\ref{tunnel} (a) is in perfect agreement with the analyses published
in Refs.~\cite{Baruselli_2015,Jacob_2015,Frank_2015}, leading to the
conclusion that the S=1 Kondo effect is actually a two-stage Kondo, where
initially a S=1/2 Kondo effect is produced by the $d_{z^2}$-orbital charge
fluctuations and the remaining magnetic moment gets screened at lower
temperatures, driven by the charge fluctuations of the $d_{x^2-y^2}$
orbital.

The above picture is qualitatively the same as $U$ increases,
Fig.~\ref{tunnel} (c).  Although the quantitative details will strongly
vary. This is in agreement with the discussion by Baruselli \textit{et
al.}~\cite{Baruselli_2015} on the values of the computed Kondo
temperatures pointing out the many difficulties to estimate accurate
values based on DFT calculations.

At contact the picture radically changes. Figures~\ref{contact} (a)
and (c) show  the half-occupied $d_{x^2-y^2}$ orbital leads to a S=1/2
Kondo effect.  The $d_{z^2}$ orbital is not relevant for Kondo physics
anymore because it becomes completely occupied. The rest of the magnetic
moment is screened by the charge fluctuations of a mixed-valence regime
driven by the degenerate $d_{xz}$ and $d_{yx}$ orbitals.  Qualitatively,
the inclusion of $U$ does not change the discussion although the final
values will greatly differ.

The above results show that the Kondo effect of Co in contact with Cu
electrodes can be considered as a single-orbital Kondo effect, at least
for a large range of temperatures in the tunneling regime, and probably
for all temperatures at contact. In this last case, non-equilibrium
effects have been discussed before~\cite{DJ_Paula}. The main effect at
contact is the increased coupling to the electrodes given by $\Gamma$.
The intrinsically non-equilibrium effects (bias-induced decoherence and
peak splitting) are largely absent from the conductance behavior in the
contact regime~\cite{DJ_Paula}.

The effect of spin-orbit interactions in Kondo processes has been
much debated in the literature.  Most works refer to the influence
of a Rashba-like spin-orbit interaction on the Kondo spin-flip
processes. The debate was very much calmed by the work of Meir and
Wingreen~\cite{Meir_1994} showing that due to the preservation
of time-reversal symmetry by the spin-orbit interaction, Kramers
degeneracy is maintained and the Kondo processes are not affected.
Recent works actually show that Rashba effects can change the
Kondo temperature reducing it~\cite{ZitKo_2011,Yanagisawa_2012} or
increasing it~\cite{Zarea_2012} depending on the system.  

Indeed,
the effect of the environment is very important.  \'Ujs\'aghy and
co-workers~\cite{Zawadowski_1996,Zawadowski_1998} showed that SOC can lead
to sizeable magnetic anisotropies depending on the environment.
This has important consequences for local spins larger than 1/2, because
it reduces the spin degeneracy and prevents Kondo spin-flip processes.
Here, we are considering the local SOC of a single impurity and not the
extended Rashba-like interactions. Since the mean-field spin of cobalt
is close to 1, we expect to find the disruptions caused by an emerging
anisotropy due to SOC and the environment of the Co atom.
However,
our calculation yields a very small magnetic anisotropy energy (MAE) when a Co atom is adsorbed
on the Cu (100) surface. The MAE is~$\sim 2$~meV. Dividing by the Boltzmann constant
yields a MAE $\sim 23$ K much smaller than $T_K\sim 90$ K, the
Kondo temperature of Co on Cu (100)~\cite{Choi_2012,Vitali_2008,Richard}.
Hence, we do not expect any effect of the SOC in the tunneling regime.
When the tip contacts the impurity, we find that the symmetry of environment of the Co atom
increases, further reducing MAE to~$\sim 0.03$ meV and unaffecting the Kondo physics.

\section{Conclusions}

Electron transport through a Co atom between an STM tip and a Cu (100) substrate
is shown to be largely independent of the Hubbard $U$ values used
in the evaluation of the electronic structure.
Despite the dramatic effects of the inclusion of correlation, transport
at the Fermi energy is basically controlled by the same orbitals. When
the tip is far from the substrate, the tunneling regime is led by
cobalt's
$sp$-electronic structure. At contact, the $d$-electronic structure
controls all electronic transport properties. Surprisingly, the electronic
transmission with or without Hubbard $U$ for both 
transport regimes is qualitatively the same and to a large
extent also quantitatively.

The inclusion of spin-orbit coupling (SOC) does not change the quantitative
values of transmission. We show that this is due to the small $\lambda/\Gamma$
ratio, where $\lambda$ is the SOC values and $\Gamma$ is the Co electronic
coupling to the electrodes. Transport with spin-orbit interactions gives
rise to spin-flip processes. At lowest-order in the above ratio we
find that the spin-flip component of the transmission 
is $T_{\uparrow,\downarrow} (E)\sim - {\lambda^2}/{\Gamma^2}$. Hence, for
Co in metals this value is very small, but for heavier impurities
will lead to sizeable decreases of electron transmission.

The effect of the Hubbard $U$ in the spin-orbit induced spin-flip transmission
is dramatic. This is due to the shifting of critical 
orbitals to be able to complete a spin-flip process. Indeed
the non-zero matrix elements of the SOC involve $\Delta S_z=\pm1$ and
$\Delta m=\pm1$ states, where $S_z$ and $m$ correspond to the  spin and
orbital-angular moment. When the values of $U_{eff}$ are
ramped from zero to 3 eV, the $\Delta m=\pm1$ states effectively
split, reducing in two orders of magnitude the spin-flip
component of the electron transmission, $T_{\uparrow,\downarrow}$.

The Kondo effect is strongly affected by the values of the
Hubbard $U$ although the qualitative picture gleaned in previous
works~\cite{Baruselli_2015,DJ_Paula,Jacob_2015,Frank_2015} remains
unchanged. Moreover, we find that the Co SOC  leads to small magnetic
anisotropy energies, well below the typical Kondo temperatures, bearing no
effect on Kondo processes for any of the conductance regimes analyzed
here.

\begin{acknowledgement}
We thank Jaime Ferrer, Pablo Rivero and Salva Barraza for
providing us with their cobalt pseudopotential. We further
thank Jaime Ferrer for providing us with a copy of Gollum
and for instructing us on its use.
We thank Roberto Robles for
many discussions and ideas.
We gratefully acknowledge support 
from MINECO (Grant No. MAT2015-66888-C3-2-R),
FEDER funds,
the CCT-Rosario Computational Center and CONICET.
\end{acknowledgement}

\bibliography{referencias}

\end{document}